\documentclass[12pt,draftclsnofoot, onecolumn]{IEEEtran}
\usepackage{balance}
\usepackage{cite}
\usepackage{graphicx}
\usepackage{psfrag}
\usepackage{subfigure}
\usepackage{url}
\usepackage{stfloats}
\usepackage{amsmath}
\usepackage{array}
\usepackage{amssymb}
\usepackage{setspace}
\usepackage{algorithmicx}
\usepackage[ruled]{algorithm}
\usepackage{algpseudocode}
\usepackage{algpascal}
\usepackage{algc}
\usepackage{color}
\usepackage{balance}
\usepackage{cite}
\usepackage{graphicx}
\usepackage{psfrag}
\usepackage{subfigure}
\usepackage{url}
\usepackage{stfloats}
\usepackage{amsmath}
\usepackage{array}
\usepackage{amssymb}
\usepackage{setspace}
\usepackage{algorithmicx}
\usepackage[ruled]{algorithm}
\usepackage{algpseudocode}
\usepackage{algpascal}
\usepackage{algc}
\usepackage{color}
\usepackage{booktabs}
\usepackage{float}

\begin{document}
\markboth{} {Duan}
\title{5G Technologies Based Remote E-Health: Architecture, Applications, and Solutions\vspace{1cm}}
\author{\normalsize Wei Duan$^{\dag}$, Yancheng Ji$^{\dag}$, Yan Zhang$^{\S}$, Guoan Zhang$^{\dag}$,
Valerio Frascolla$^{\ddag}$ and Xin Li$^{\flat}$\\
\vspace{1cm}$^{\dag}$School of Information Science and Technology,\\
Nantong University,
Nantong 226000, China\\
$^{\S}$Maternity and Child Care Hospital, Edong Medical Group, Huangshi 435000, China\\
$^{\ddag}$Division of Research and Innovation, Intel Corporation\\
$^{\flat}$Department of Physical Education,\\
Zhengzhou University, Zhengzhou 450000, China\\
Email: sinder@ntu.edu.cn, jiyancheng@ntu.edu.cn,
yzhang$\_$edong@hotmail.com, gzhang@ntu.edu.cn,
valerio.frascolla@intel.com, xinlibox@zzu.edu.cn} \maketitle

\begin{abstract}
Currently, many countries are facing the problems of aging
population, serious imbalance of medical resources supply and
demand, as well as uneven geographical distribution, resulting in
a huge demand for remote e-health. Particularly, with invasions of
COVID-19, the health of people and even social stability have been
challenged unprecedentedly. To contribute to these urgent
problems, this article proposes a general architecture of the
remote e-health, where the city hospital provides the technical
supports and services for remote hospitals. Meanwhile, 5G
technologies supported telemedicine is introduced to satisfy the
high-speed transmission of massive multimedia medical data, and
further realize the sharing of medical resources. Moreover, to
turn passivity into initiative to prevent COVID-19, a broad area
epidemic prevention and control scheme is also investigated,
especially for the remote areas. We discuss their principles and
key features, and foresee the challenges, opportunities, and
future research trends. Finally, a node value and content
popularity based caching strategy is introduced to provide a
preliminary solution of the massive data storage and low-latency
transmission.

\end{abstract}

\section{Introduction}

With the extreme unbalanced distribution of medical resources,
there is a big gap between the developed areas and economically
backward areas in terms of the equipment, technology service
quality of medical, resulting in rapid demands for telemedicine
\cite{R3}. The original intention of telemedicine is to improve
the popularity of medical and health services via
telecommunication for medics \cite{R2}. With the strong support of
market policy and progress of wireless technology, telemedicine
has been developed significantly \cite{R4}. Currently, relying on
the advanced communication and computer technologies to transmit
the data, voice, image, video and other information, telemedicine
can realize the treatment, diagnosis, health care and consultation
in real-time for the remote patients, as well as provide the
education and training for remote medics, which breaks the space
and time limitations \cite{R4, R5}. Moreover, the telemedicine not
only changes the medical experience for patients, but also
improves the medic-patient relationship. When the patients seek
medical treatment, the medic will take their emotions into account
to strive for positive treatment evaluations. It is easy to see
that, telemedicine will break the barriers among different
industries, optimize the medical service process, improve the
overall service efficiency, and constantly resolve the problems
provided by complicated medical procedures.

As the core support of telemedicine, with decades of development
and continuous consumption upgrading, the wireless communication
technology has completed the evolutions from 1G to 5G \cite{o1,
o2,o5}. It realizes the high-quality transmission of three
dimensional images to provide high-quality video services，data
acquisition, positioning, remote diagnosis and treatment and other
fusion functions in real-time. Compared with other generations of
wireless communications, 5G has advantages in terms of the low
latency, high reliability and mobility, providing great
opportunity for the development of telemedicine \cite{o3}. On the
basis of traditional medicine, 5G technologies based telemedicine
integrates mobile communication, Internet, Internet of things
(IoT) \cite{o2}, cloud computing, big data, artificial
intelligence (AI) \cite{o4} and other advanced information and
communication technologies, applying to the remote surgery, remote
consultation, remote health monitoring and emergency command. In
particular, telemedicine will provide more choices and ways for
rescue, especially in the fast moving state of vehicle and harsh
environment.

It worth noting that, since that the 5G technology, business model
and industrial ecology are still evolving and exploring, the
architecture, system design and landing mode of telemedicine are
not completed. These arise the following problems problems: The
imperfect overall planning, and the problem of cross departmental
coordination; lack of technical verification and feasibility
study; inconsistent medical standards; privacy security \cite{o5,
o7}. On the other hand, with the spread of COVID-19 \cite{R1, o6},
physical and mental health of people has been greatly impacted,
leading to that the concern of people has gradually transferred
from the disease treatment to disease prevention and health
management. Moreover, in order to realize remote sharing of
medical resources, the massive data storage and data redundancy
will bring great load to the server. With these observations, the
goal of this article is to provide a potential solution to realize
5G technologies-based remote e-health, spanning from the general
architecture and framework of telemedicine, to satisfy the
high-speed transmission of massive multimedia medical data and
realize the sharing of medical resources. In order to track and
control the spread of the COVID-19, the broad area epidemic
prevention and control (BAEPC) design for COVID-19 is proposed, as
well as the node value and content popularity (NVCP) based caching
strategy is investigated to overcome the massive data storage and
low-latency transmission issues.

The rest of this article is organized as follows. First, we
provide a general architecture of the remote e-health. Then the 5G
technologies based telemedicine framework is introduced for the
remote hospital. Moreover, a broad area epidemic prevention and
control scheme is investigated to prevent COVID-19, as well as the
node value and content popularity based caching strategy is
studied. Finally, we draw the main conclusions and interesting
future research.

\section{The proposed Remote E-Health Architecture}

\begin{figure}
\begin{center}
\includegraphics [width=160mm]{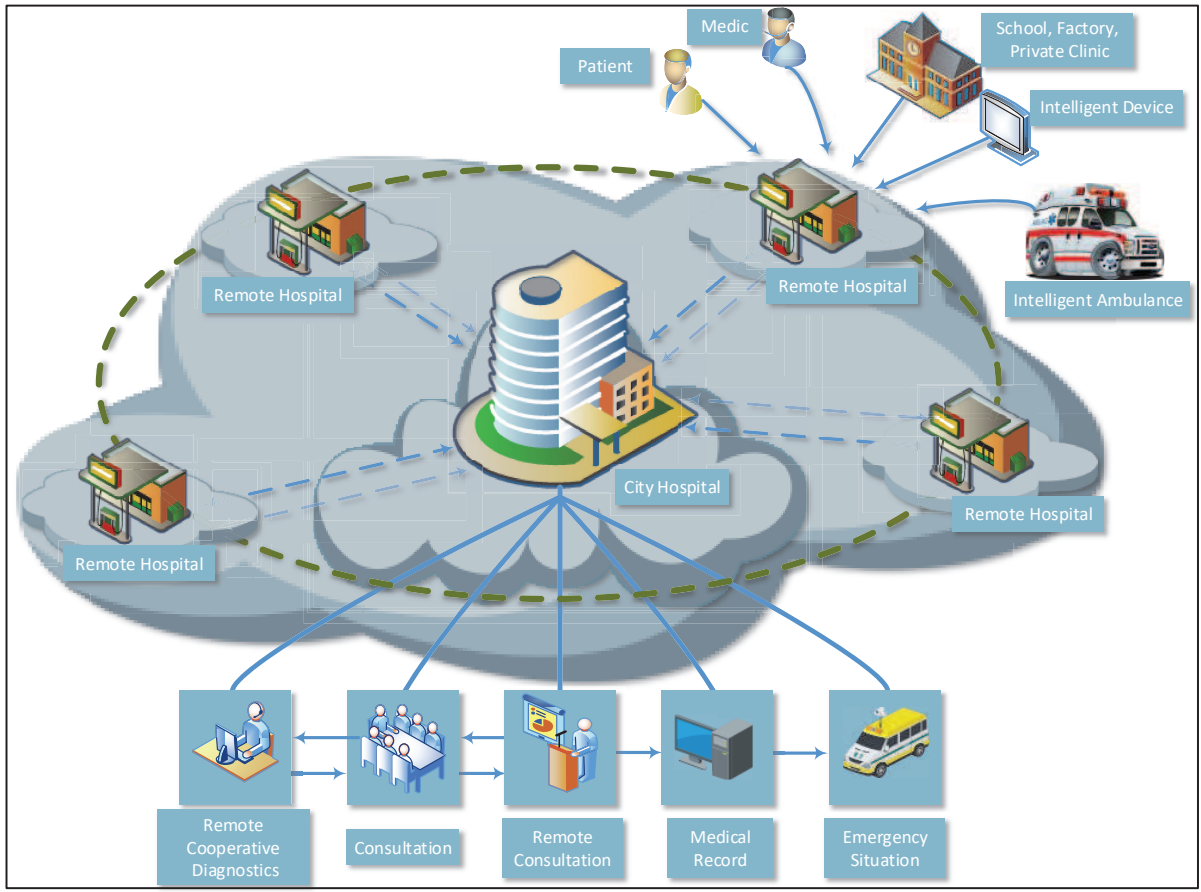}\\
\caption{\label{9} Illustration of the remote e-health
architecture.}
\end{center}
\end{figure}

Relying on computer technology and remote sensing, telemetry,
remote control technologies, telemedicine gives play to advantages
of medical technologies and equipments in city hospital to conduct
remote diagnosis, treatment and consultation for patients in
remote areas，i.e., remote imaging, remote nursing and other
medical activities. The proposed remote e-health architecture
based on cloud network is shown in Fig. 1, which consists of the
city hospital and many corresponding remote hospitals. The
concepts of the proposed architecture is that, with the internet
as link, grading diagnosis and treatment as the core and the
substance hospital as the support, the remote hospitals and
advanced city hospitals will be connected to this platform. By
this way, the remote hospitals can also enjoy the remote
outpatient service, expert appointment, electronic prescription,
online payment and other fast services through the internet. As
the brain, the city hospital provides the technical supports and
services for these remote hospitals, in the meanwhile that the
remote hospitals share information and data for each other
according to the networks, to improve the utilization of medical
resources. For the city hospital, the details of the processing
strategies can be summarized as follows:
\begin{itemize}
\item When a request for medical help from a remote hospital is
received, according to the received contents, i.e., the images,
voices and videos for the patients, the city hospital rapidly
makes decisions and corresponding measures to cooperatively help
remote hospital curing the patients, through the existing advanced
technologies and equipments. \item For the difficult miscellaneous
diseases, the city hospital convenes experts and relevant medics
to hold the consultation. Moreover, for very special and difficult
cases, the remote consultation with other advanced city hospitals
will be adopted. When the specific treatment plan is formulated,
the city hospital will promptly contact and assist the remote
hospital to take corresponding measures. In the meanwhile, the
electronic medical record is established. \item According to the
progress of conditions of the patients, the electronic medical
record will be updated in real-time, until the patient is fully
recovered. The electronic medical records are also shared with the
remote hospitals for follow-up actions and future study. Moreover,
for emergencies, the city hospital will dispatch the intelligent
ambulance and medics to the remote hospitals.
\end{itemize}
All the city and remote hospitals will share and update the
information through the cloud network. Clearly, the use of
telemedicine not only significantly reduce the time and cost of
the diagnosis and treatment, but also can well manage and
distribute emergency medical services in remote areas.
Specifically, it can make medics break through the limitation of
geographical scope and share the case and diagnosis photos of
patients, which is conducive to the development of clinical
research. In addition, it can provide a better medical education
for medics in remote areas.

Since that the telemedicine technology is in its development
stage, the design of its architecture and corresponding strategies
are different from the traditional medical system. The key issues
and challenges for telemedicine are generally summarized as
follows:
\begin{itemize}
\item \textbf{Privacy security}: Any breakthrough in science and
technology has to face the problem of security, the telemedicine
technology is no exception. If the medics or medical equipments do
not consider the security of electronic data of patients, once
these data are transmitted and leaked through the Internet, it
will cause irreparable security risks. Therefore, it is necessary
that, adopting 5G technology and network security methods to
authenticate, encrypt and protect the intelligent medical
equipment for privacy preservations. Only by taking precautions in
advance, remote medical can realize the transformation from the
passive defense to active response. \item \textbf{Medical data and
resource sharing}: Medical data and resource sharing can not only
help the rapid development of the telemedicine technology, but
also significantly alleviate the shortage of medics. However, when
telemedicine is performed, it has to connect to Internet, and in
this docking process, the systems of hospitals are relatively
closed; the electronic systems of different hospitals are built by
different enterprises; and there exists barriers between these
systems among enterprises, resulting in a difficult integration
for the data from different hospitals. Therefore, how to
reasonably and legally realize the sharing of massive medical data
to the Internet is still an open problem and challenge. \item
\textbf{Massive connectivity and data cache}: With the commercial
application of 5G, the real-time data transmission problem for
telemedicine technology has been solved in some degree,
eliminating the barriers and distance for medical communication.
However, the massive connectivity from the medical devices,
intelligent devices and remote hospitals, as well as the cache of
the massive medical data challenges the existing spectrum
resources and network structure. Therefore, it is necessary to
adopt the technologies with the excellent spectrum efficiency and
effective cache capacity.
\end{itemize}

\section{5G Technologies Based
Telemedicine Framework}

\begin{figure}
\begin{center}
\includegraphics [width=160mm]{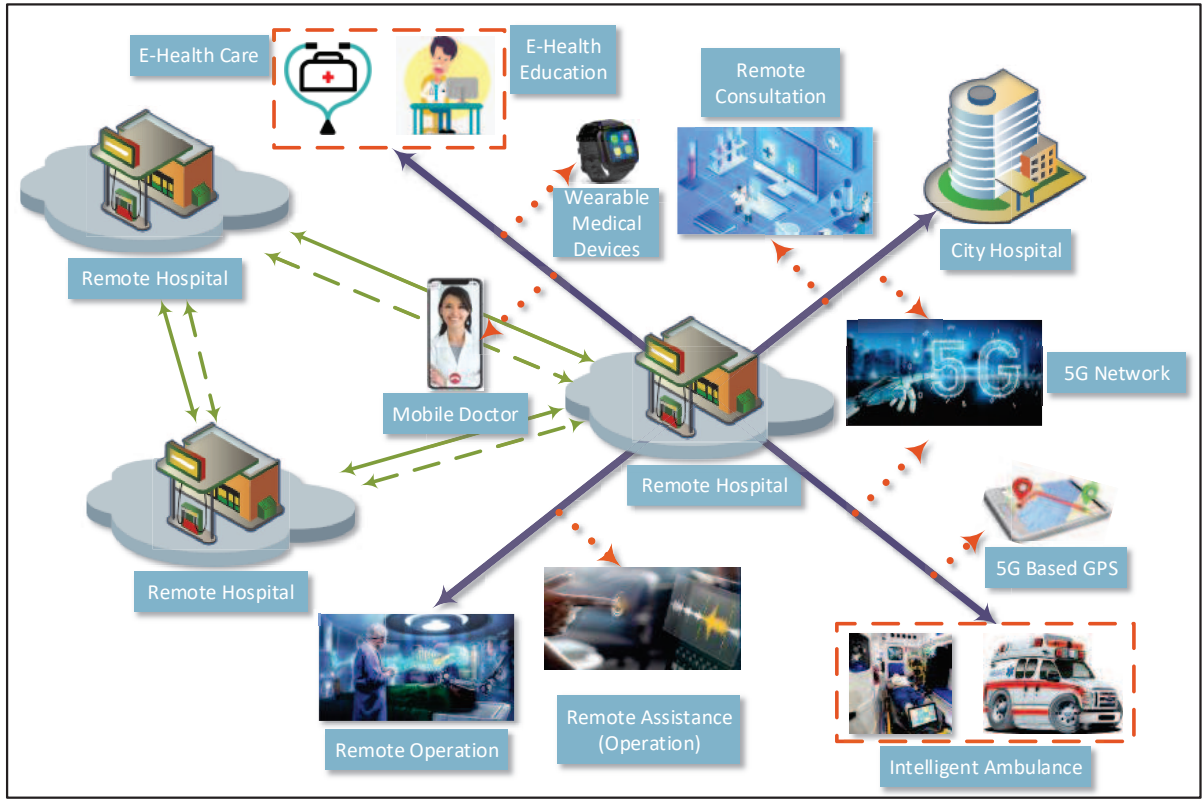}\\
\caption{\label{9} 5G technologies based telemedicine framework.}
\end{center}
\end{figure}

On the basis of traditional medicine, 5G technologies based
telemedicine integrates wireless communication technology of smart
equipment and high-speed mobile communication technology in
various modes, which can realize the operation of remote surgery,
remote consultation, patient monitoring, command and
decision-making for emergency rescue events. Moreover, 5G-based
telemedicine can also support the high-speed transmission of
massive multimedia medical data, and further realize the sharing
of medical resources. With this prospect, as shown in Fig. 2, the
remote hospital is readily allowed the patients, local medics,
schools, factories, personal devices and local intelligent
ambulances access to its server to apply the medical resources and
share the medical data.

Nowadays, medical service has changed from the disease treatment
to health care, meanwhile, the disease prevention and health
management are becoming increasingly important. With the wearable
medical devices and mobile private doctor, people can know their
personal physical signs, i.e., blood pressure, heart rate and
temperature, at any time and any where to enjoy high quality
health services and e-health education. In addition, through the
monitoring of these devices, medical institutions and medics can
take the initiative to find individuals and groups with abnormal
health status, and give health risk tips, health improvement or
medical measures suggestions in advance. In this manner, the
hospitals can improve diagnosis efficiency, and residents can
reduce the cost of health consultation. In addition, based on
internet of medical things (IoMT) and AI, for any emergency, the
patients can be timely and tentatively cured in the ambulance to
realize the vision of ``In ambulance, in hospital". According to
the 5G HD video feedback from the ambulance, the hospital can
conduct real-time follow up and analyze the signs and conditions
of patients in advance, to effectively reduce the risk of death.
On the other hand, with the development of 5G-based global
positioning system (5G-GPS), it can provide more accurate
positioning, more intelligent navigation and more information
services in real time for the patients and ambulance, especially
for remote areas. Predictably, telemedicine can improve the
medical experience of the patients, and constantly resolve the
problems of ``complicated treatment process". Moreover, it also
provides more possibilities to make up for the insufficient and
unbalanced distribution of medical resources and solve the problem
of social aging.

\section{Broad Area Epidemic Prevention and Control for COVID-19}

\begin{figure}
\begin{center}
\includegraphics [width=160mm]{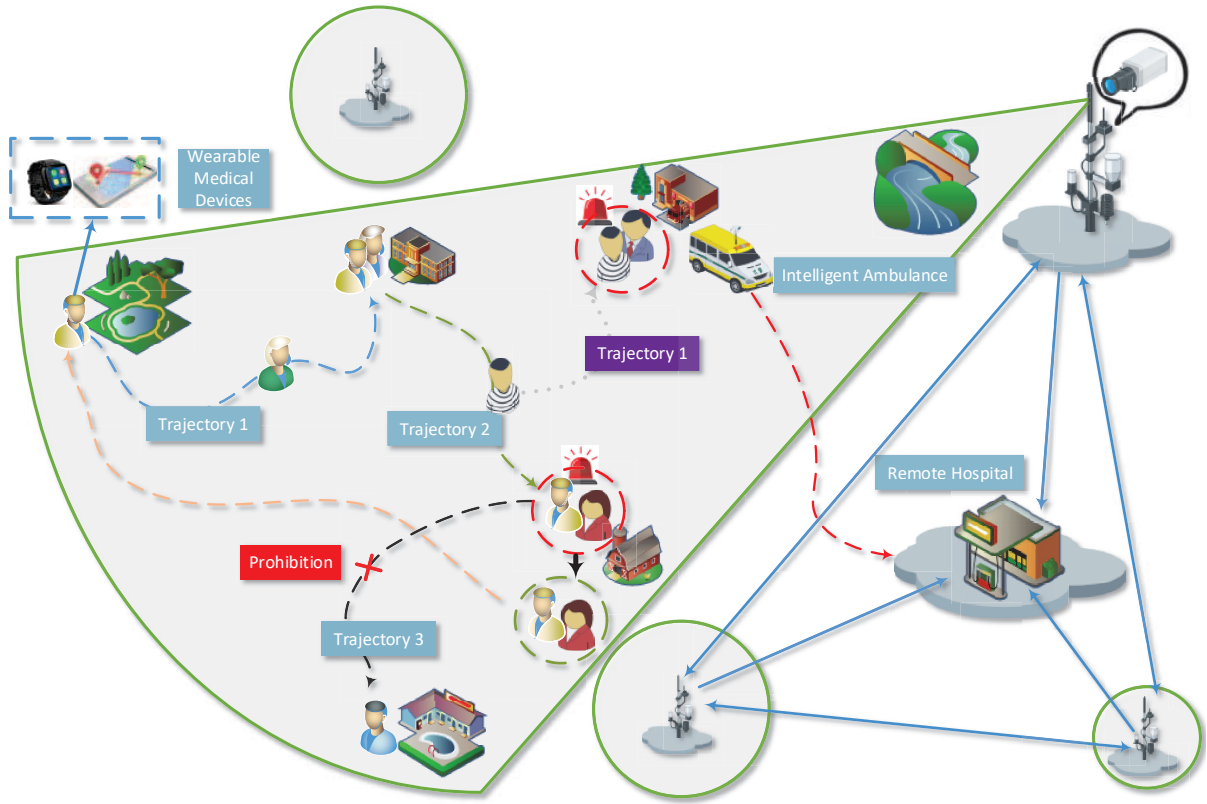}\\
\caption{\label{9} The broad area epidemic prevention and control
scheme.}
\end{center}
\end{figure}

With invasions of COVID-19, due to the continuous person-to-person
transmission, the coronavirus rapidly spreads leading to cross
infection for many patients. Since that there is no effective cure
method and vaccine, and it is hard to detect millions of people on
a large scale, the strict segregation and control measures have to
be adopted. Unavoidably, the economic development and quality of
life of the people have been greatly impacted, even resulting in a
social panic. Without radical cure, effective and rapid detection
to prevent the spread of the coronavirus has become the primary
task. Currently, the common detection method is that, at the
entrance and exit with large flow of people, the thermal cameras
or temperature guns are used to locally detect the temperature of
people in turn. Clearly, such detections have the following
defects:
\begin{itemize}
\item \textbf{Omissions in personnel inspection}: The tested
personnel are passively restricted, not all of them will be
detected. For example, some people do not take the initiative or
cooperate with the measurement, especially that the people in
remote areas have weak awareness of protection; \item
\textbf{Real-time issue}: This kind of epidemic prevention and
control is not real-time due to the rapidly spreading of
coronavirus. It is inevitable to cross infection in the detection
process, especially in remote areas; \item\textbf{Locality issue}:
Due to that COVID-19 is a global problem, it is difficult to make
personnel information open and personnel information transparent
among different regions, which makes it necessary to provide a lot
of manpower and material resources when people flow between
regions; \item \textbf{Security issue}: On one hand, patient
information is presented by the text registration; on the other
hand, most of the body temperature and pathological features are
shown in the form of pictures. It is easy to see that this
intuitive way will inevitably be used by eavesdroppers providing
troubles to patients.
\end{itemize}
In order to turn passivity into initiative, a BAEPC for COVID-19
is proposed as shown in Fig. 3. With the development of the
high-definition cameras and video surveillance, currently, ultra
long distance thermal camera (ULDTC) can monitor a circumference
of $15$ Km. The basic idea of this scheme is that distribute these
rotatable ULDTC in different areas for independent monitoring, and
centralize the collected information to the control center (remote
hospital) via the 5G-network for centralized processing. In
addition, the people should carry wearable medical devices, by
this way, the trajectories of people will be collected by the
remote hospital to determine coordinates of people during their
outdoor activities. In this manner, the people can receive
personal information and surrounding conditions from the remote
hospital at any time, to avoid cross infection when abnormal body
temperature occurs. Accordingly, when people themselves or close
contacts have abnormal body temperature, they will receive warning
messages in time and make self isolation until temperature normal
or $14$ days. Due to huge amount of data, it is considered that
the people staying at home or in their vehicles are isolated, the
remote hospital will not collect their coordinates until they go
out for activities or take the initiative to contact remote
hospital. When the fever have stayed high, after receiving the
request for help, the patient will be sent to the remote hospital
for a further observation and treatment by the ambulance.

\section{Node Value and Content Popularity Based Caching Strategy}

Even that, the proposed BAEPC scheme can effectively and promptly
confine and eliminate the coronavirus, however, the massive data
storage and data redundancy will bring great load to the server.
Moreover, due to that the key of telemedicine technology lies in
long-distance and low-latency connections, TCP/IP networking
approach is hard to satisfy these requirements. In this section, a
NVCP based caching strategy for content-centric networking (CCN)
will be introduced to provide a preliminary solution. In what
following, after defining the cache content, the proposed NVCP
caching strategy will be discussed within two algorithms.

\subsection{Cache locality}

In this subsection, three node attributes are defined to evaluate
the value of node, which are based on the graph theory and
described. Moreover, we further considered that the Named-data
Link State Routing Protocol (NLSR) is adopted to query the
shortest path information. Given an undirected graph $G=(V,E)$
with \emph{n} vertexes and \emph{m} edges, where
$V=\left\{v_{1},v_{2},...,v_{n}\right\}$ represents a set of
content routers, and $E=\left\{e_{1},e_{2},...,e_{m}\right\}$
denotes the links between the content routers. Moreover,
$A=(a_{ij})_{n\times{n}}$ is the adjacency matrix of $G$, for
$v_i$  directly connect with $v_j$ and $a_{ij}=1$, otherwise
$a_{ij}=0$.

\begin{enumerate}
\item \textbf{\emph{Connectivity}:} Different forwarding
strategies result in different routing paths for the requested
content, cache nodes will play different roles in these
strategies. And hence, we regard the number of paths that the
requested content pass through the cache node as the connectivity
of the node. Therefore, with the increasing paths, the request
content becomes more important. Defining the number of routing
paths, which is requested content \emph{k} passes through $v_i$,
as $c_s(v_i)$, and the maximum number of routing paths passing
through $v_i$ as $c_s^{max}(v_i)$, the connectivity can be
obtained as the ratio of $c_s(v_i)$ to $c_s(v_i)^\emph{max}$
defended as $C_s(v_i)$.

\item \textbf{\emph{Betweenness centrality}:} If a content router
is on the shortest paths between the corresponding content
routers, the content router is considered to be in a significant
position. It is reasonable, due to that the content router in this
position can affect the overall network by controlling or
misinterpreting the transmission of information. The ability to
characterize content router control information transfer is
betweenness centrality (also known as node median) \cite{a2}.
Defending $\sigma_{st}$ as the number of shortest paths between
$v_s$ and $v_t$, $\sigma_{st}(v_i)$ as the number of shortest
paths from $v_s$ to $v_t$ through $v_i$, the betweenness
centrality of $v_i$ can be presented as
\begin{eqnarray}\label{2}
C_{\emph{B}}(v_i)=
\left(\frac{(n-1)(n-2)}{2}\right)^{-1}\sum_{s\ne{t}\ne{i}\in{v}}\frac{\sigma_{st}(v_i)}{\sigma_{st}},\notag
\end{eqnarray}
where $n$ represents the number of content routers.

\item \textbf{\emph{Eigenvector centrality}:} In fact, the
influence of a content router is not only related to its own
locality, but also to the influence of its neighbors \cite{c18}.
If the content router is chosen by a very popular actor, the
corresponding influence will also be increased. On the other hand,
there is an influence on an influential node, it is clear that the
influence will be even greater, where the eigenvector centrality
is used to characterize the influence. We define
$C_{\emph{E}}(v_i)$ as the eigenvector centrality of a node,
indicating the influence of the neighbors of nodes. It is also
defended that $C_{\emph{E}}(v_i)$ not only reflects the relative
centrality of the network, but also reflects the long-term
influence of the node.
\end{enumerate}

The connectivity and betweenness centrality consider the value of
nodes from routing paths of the requested contents, meanwhile that
the eigenvector centrality takes the influence of neighbors into
account. When select the cache locality, the NVCP considers the
above three attributes simultaneously. Defining $M(v_i)$ as the
comprehensive attribute, we have:
\begin{eqnarray}\label{4}
   M(v_i)=\alpha C_{\emph{S}}(v_i)+\beta C_{\emph{B}}(v_i)+\gamma
   C_{\emph{E}}(v_i),\notag
\end{eqnarray}
where $\alpha, \beta, \gamma$ represent the weight of
connectivity, betweenness centrality and eigenvector centrality,
and the sum of them is $1$. It is worth noting that, in our
proposed scheme, three mentioned attributes have difference
influences on the chosen of the cache locality. Based on which,
when different attributes are used to evaluate the importance of
nodes in a same network, the corresponding different results will
be obtained. Therefore, the coefficients in the comprehensive
attribute $M(v_i)$ are determined by the related requirements of
CCN.

\subsection{Cache content}

Since that whether caching every content which pass through the
content router is another problem for the CCN, the popularity is a
factor to draw the content. The popularity of content can be
estimated by the content request count during a measurement, which
means that the more content request counts, the greater the
popularity and probability of the content will be requested.
Assuming that the count requesting for the content $\emph{k}$ at
$v_i$ is $f_{v_{i},k}$, and the max count of $v_i$ is
$f_{v_i}^\emph{max}$, finally, we have the popularity of content
$\emph{k}$ can be presented as
$P_{v_i}(k)=\frac{f_{v_i,k}}{f_{v_i}^{max}}$.

\subsection{The NVCP cache strategy}

For the proposed NVCP, the core idea is based on the node value
and content popularity, a table is considered to be added at each
content node including the content name, the number of routing
path and count of content request to store the information of
content and cache node. It is remarkable that, in CCN/NDN, PIT
records the requests that have not been satisfied, including the
content name and corresponding arrival interface, to ensure the
returned response packet to the content requester along the
reverse path. Therefore, the source of a request is identified
through PIT. By this way, when a consumer requests a content, the
betweenness centrality and eigenvector centrality of the nodes on
the delivery path will be calculated and normalized. Once the
request is satisfied, the data packet is returned on the inverse
delivery path. At this time, the content popularity will be
calculated according to the count of content request. In our
proposed scheme, we design a variable $\varphi$ to match the
content popularity and node value given as $
\varphi=\frac{P_{v_{i},k}}{M(v_i)}, $ where $P_{v_i}(k)$ is the
popularity of content $k$ at $v_i$, and the values of
$P_{v_{i},k}$ and $M(v_i)$ are fixed and less than $1$. In
general, there are two cases: (1) $P_{v_{i},k}\geq M(v_i)$, it
means that the popularity of content is more important than the
value of node. Therefore, caching the content in the content
router can obtain a higher cache hit rate. (2)
$P_{v_{i},k}<M(v_i)$, it means that the value of the node is high,
but the corresponding popularity of the content is low. If caching
content with a lower popularity will result in a waste of the
cache space.

\begin{table}
\caption{Obtain the betweenness centrality and eigenvector
centrality}
\begin{center}
\begin{tabular}{l}
\toprule  
\textbf{Algorithm 1:}\quad Set the forward path   \\
\midrule  
\textbf {G:} The network topology  \\
\textbf{Initialize} $c_S(v_i), C_B(v_i), C_E(v_i), f_{v_{i},k} $    \\
\textbf{for} node on the delivery path from consumer to sever \\ \textbf{do} \\
\hspace*{5 mm}\textbf{if} content in cache \\
\hspace*{8 mm}\textbf{then} send content back to the consumer \\
\hspace*{16 mm}discard interest packet \\
\hspace*{5 mm}\textbf{else} \\
\hspace*{8 mm}\textbf{get} the adjacency matrix of the nodes according G \\
\hspace*{8 mm}$\sigma_{st}$: record the number of shortest paths between \\
\hspace*{15 mm}$v_s$ and $v_t$ \\
\hspace*{8 mm}$\sigma_{st}(v_i)$: record number of shortest paths from $v_s$ \\
\hspace*{21 mm}to $v_t$ through $v_i$ \\
\hspace*{8 mm}\textbf{calculate}\quad $C_B(v_i), C_E(v_i)$ \\
\hspace*{8 mm}$c_S(v_i)\leftarrow c_S(v_i)+1$ \\
\hspace*{8 mm}$f_{v_{i},k}\leftarrow f_{v_{i},k}+1$ \\
\hspace*{7 mm} forward the interest packet to the next hop towards \\
\hspace*{7 mm} server \\
\hspace*{5 mm}\textbf{end if}\\
\textbf{end for}\\
\bottomrule 
\end{tabular}
\end{center}
\end{table}

\begin{table}
\caption{Select the appropriate cache locality and cache content}
\begin{center}
\begin{tabular}{l}
\toprule  
\textbf{Algorithm 2:}\quad Select cache locality and cache content  \\
\midrule  
\textbf {G:} The network topology  \\
\textbf{Input} $c_S(v_i), C_B(v_i), C_E(v_i), f_{v_{i},k} $ \\
\textbf{for} node on the delivery path from server to consumer \textbf{do} \\
\hspace*{5 mm}\textbf{if} the content is provided by server \\
\hspace*{9 mm}\textbf{then} send the data packet back directly \\
\hspace*{5 mm}\textbf{else} \\
\hspace*{10 mm}\textbf{calculate}\quad $C_S(v_i), P_{v_{i},k}$ \\
\hspace*{10 mm}\textbf{get} $C_B(v_i), C_E(v_i)$ \\
\hspace*{10 mm}$M_{v_i}\leftarrow\alpha C_{\emph{S}}(v_i)+\beta C_{\emph{B}}(v_i)+\gamma C_{\emph{E}}(v_i)$ \\
\hspace*{5 mm}\textbf{end if} \\
\hspace*{5 mm}\textbf{if} \quad $\varphi=\frac{P_{v_{i},k}}{M(v_i)}\ge1 $\\
\hspace*{9 mm}\textbf{then} cache the contents \\
\hspace*{5 mm}\textbf{else} \\
\hspace*{7 mm} forward the data packet to the next hop to the \\
\hspace*{7 mm} consumer \\
\hspace*{5 mm}\textbf{end if} \\
\textbf{end for}\\
\bottomrule 
\end{tabular}
\end{center}
\end{table}

The main idea of the proposed NVCP is presented in Algorithms 1
and 2. In our proposed scheme, considering that the location of
content router does not change, we have a fixed network topology.
Therefore, the network can be seen as an undirected graph, the
corresponding algorithms (such as Brande algorithm and Power
Iteration) will be used to obtain $C_B(v_i)$ and $C_E(v_i)$ in
advance, resulting in a computational complexity as $\mathcal
O(VE)$ for these two algorithm. Algorithm 1 is the process to
obtain the betweenness centrality and eigenvector centrality. It
is clear that, when the interest packet arrives at a content
router, if the CS has the content, sends the content back to the
consumer, otherwise calculates $C_B(v_i)$ and $C_E(v_i)$ according
to the network topology. In the meanwhile, the values of
$C_S(v_i)$ and $f_{v_i,k}$ increase by $1$. On the other hand,
algorithm 2 illustrates the process to select the appropriate
cache locality and cache content. According to the results given
in Algorithm 1, calculate $\varphi$. If $\varphi>1$, cache the
content, otherwise forward the data packet to the next hop. In
addition, considering the fixed locations of content routers, the
values of $C_B(V_i)$ and $C_E(V_i)$ only need to be calculated
once. By this way, when be requested, the popularity of content
increases by $1$, which is easy to realize. Clearly, compared with
the existing works, our proposed algorithm significantly improve
the efficiency for calculating the value of $\varphi$. Clearly,
the computational complexities of Algorithms $1$ and $2$ are not
extremely high, which are practical and acceptable.

\subsection{Simulation Results}

The simulation uses a network topology generated randomly, which
consists of $50$ nodes and $150$ links. There is a source server
in the network, which is connected to a node randomly, and the
edge nodes are connected to the consumers. Content requests are
generated following the Zipf-Mandelbrot distribution with $a=0.7$.
The total number of different contents will be requested in the
network as $10,000$. Further assume that the interests of each
consumer are generated following the Poisson distribution with
$\lambda=100/s$. Comprehensive consideration of the various
attributes of the node, for simplicity and fairness, in this
article, the specific weight values of $\alpha$ (connectivity),
$\beta$ (betweenness centrality), and $\gamma$ (eigenvector
centrality) in the presented simulation results are equivalently
given as $1/3$. The Least Recently Used (LRU) \cite{a3} is
employed as the cache replacement strategy and the total
simulation time is $100$s. More specially, the simulations results
have been evaluated for various values of the cache size. The main
simulation parameters are listed in Table III.

\begin{table}[!htbp]
\centering \caption{Simulation parameters}
\begin{tabular}{ccc}
\toprule  
Parameter & Default value & Variation range\\
\midrule  
Nodes & 50 & -\\
Links & 150 & -\\
Delay/ms & 10 & -\\
Bandwidth/Mbps & 10 & -\\
Contents & 10,000 & -\\
Consumers & 18 & -\\
Cache size & 1,000 & $100\sim2,000$ \\
zipf(a) & 0.7 & $0.1\sim1.0$ \\
Simulation time/s & 100 & - \\
\bottomrule 
\end{tabular}
\end{table}

\begin{figure}
\begin{center}
\hspace{-8mm}\includegraphics[width=0.3725\textwidth]{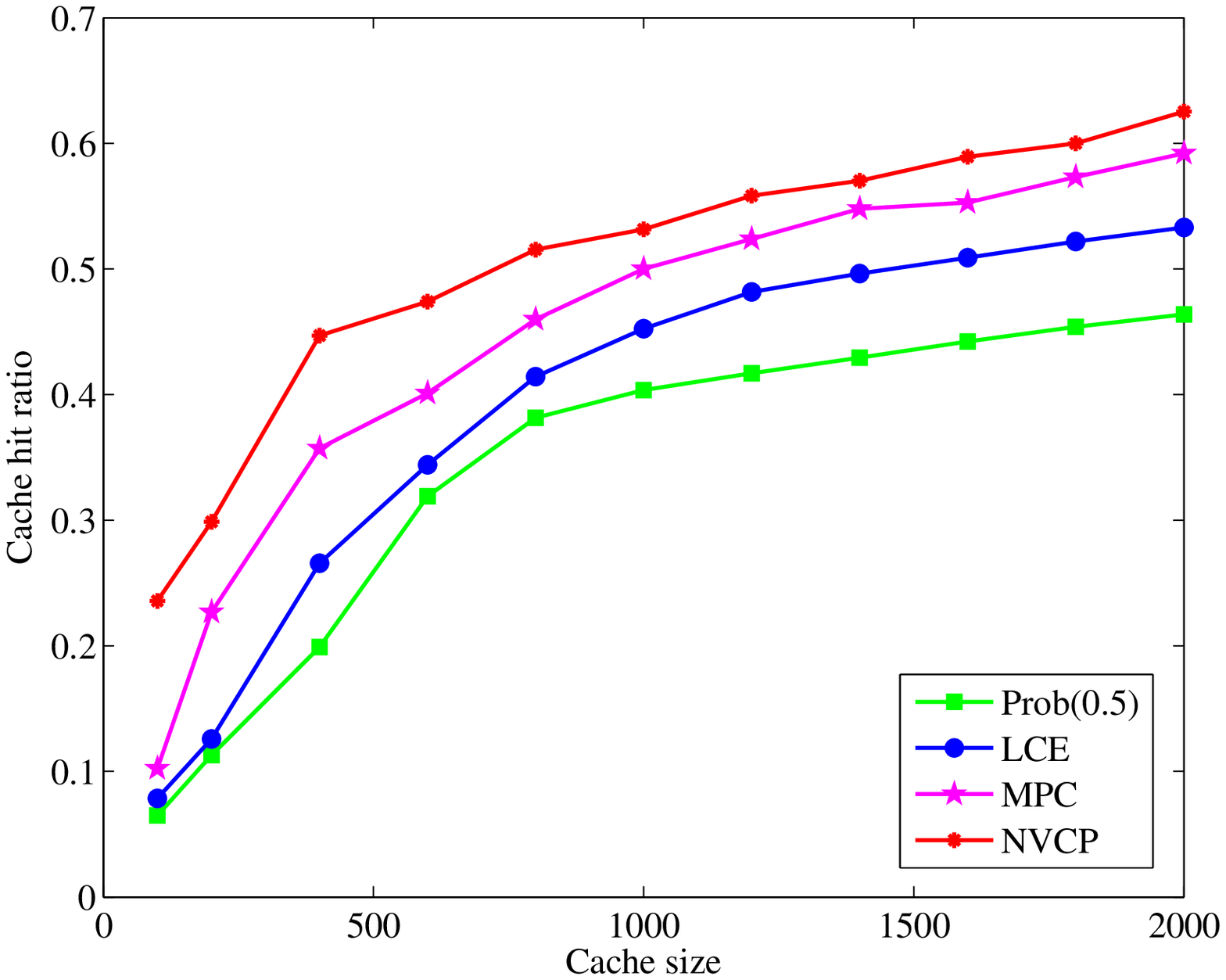}
\hspace{-7mm}\includegraphics[width=0.3725\textwidth]{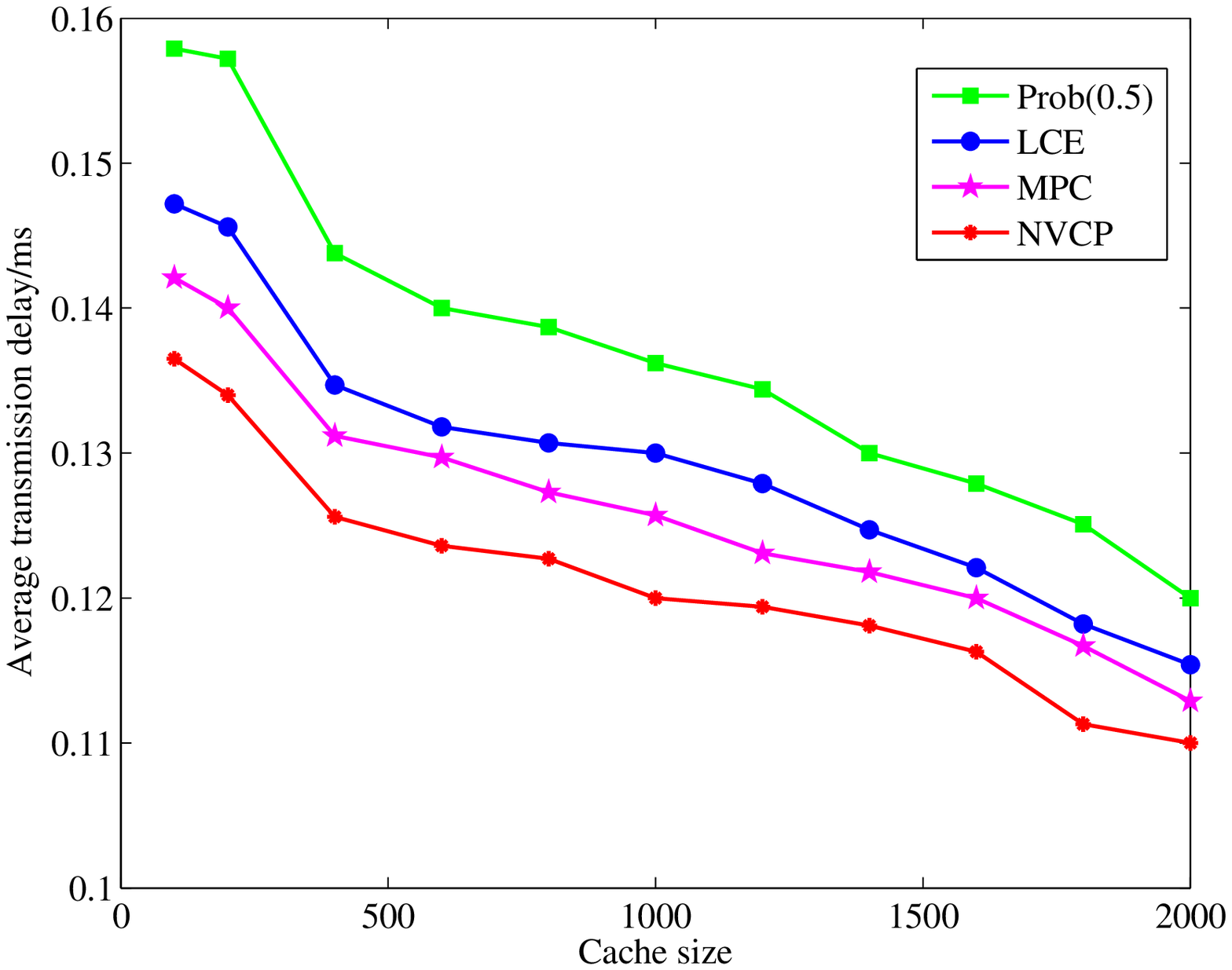}
\hspace{-7mm}\includegraphics[width=0.3725\textwidth]{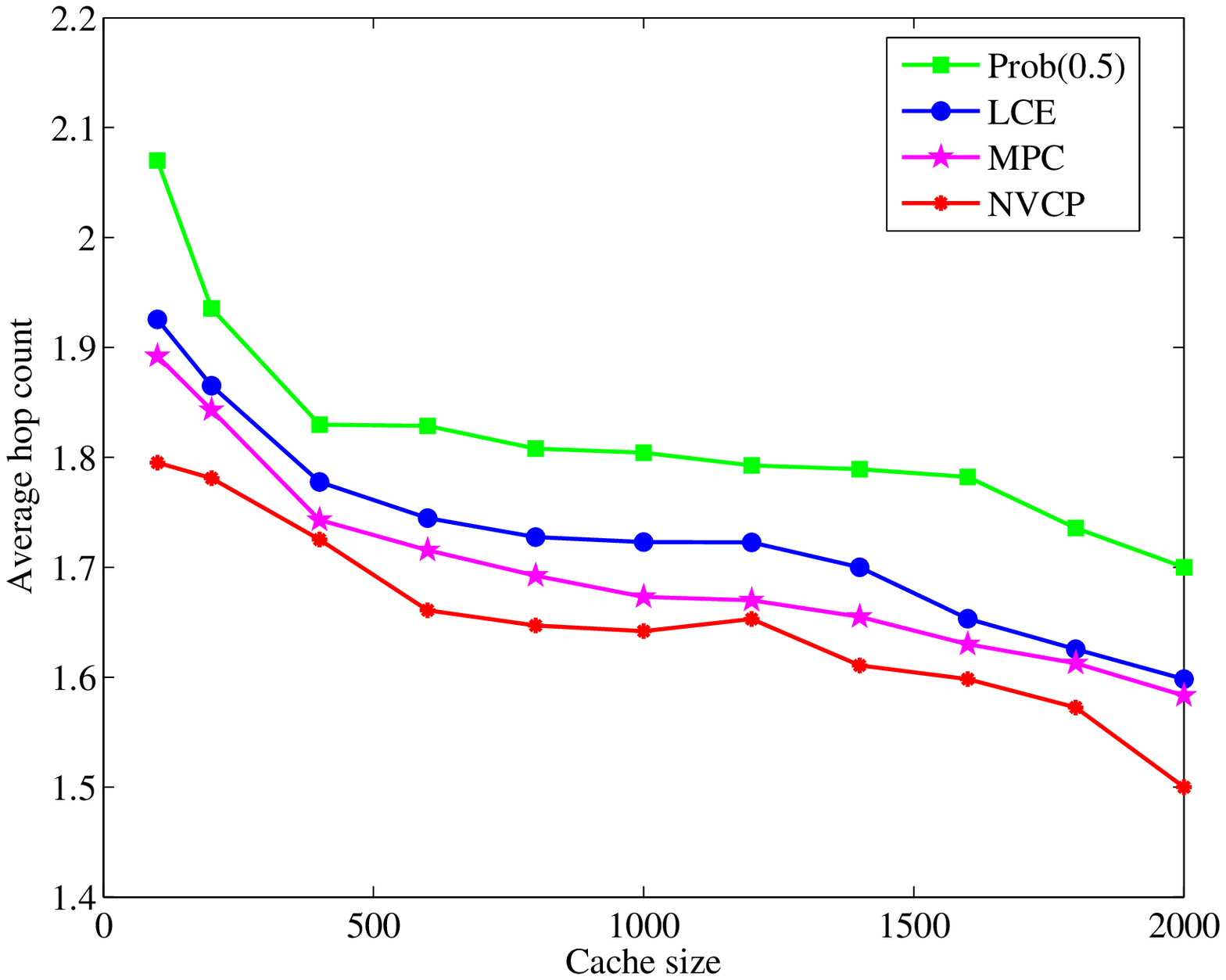}
\caption{The impact of cache size on the system performance for
the proposed and existing caching schemes versus the cache size.}
\end{center}
\end{figure}

The proposed NVCP strategy is compared with the LCE, Prob(0.5) and
MPC in terms of the cache hit ratio, average hop count and average
transmission latency as show in Fig. 4. It is easy to see that the
cache hit ratios of the four cache strategies are gradually
increased, and the cache hit ratio of the NVCP is significantly
better than the others. It is resealable, because the LCE requires
all nodes on the delivery path cache contents without difference,
which results in a large amount of content redundancy and replace
frequently. In addition, the Prob(0.5) caches contents passing
through the cache nodes with a fixed probability. Even taht the
cache space is reduced, it still causes the content redundancy and
low content diversity. Instead of storing all the content at each
node on the path, MPC caches only the popular contents. On the
contrary, the NVCP considers node value and content popularity
comprehensively, where the content with higher popularity is
cached in nodes with higher value, in the meanwhile that the
content with lower popularity is cached in nodes with lower value,
which significantly reduces the replacement frequency, improves
the content diversity, and reduces the content redundancy.
Compared to the LCE, Prob(0.5) and MPC schemes, the proposed NVCP
cache hit rate has a $11\%$ to $15\%$ improvement. The second and
third subfigures show that as the cache capacity of the node
increases, the average hop count and the average transmission
delay decrease gradually. Moreover, the performance of NVCP is
better than the other schemes. This is due to that the LCE caches
content indiscriminately, Prob(0.5) takes the probability caching,
and the MPC only caches the most popular content without any
requirements for the nodes. On the contrary, the NVCP
comprehensively evaluates node value from the connectivity,
betweenness centrality and eigenvector centrality, assigns
different weights according to different requirements, which
improves the response speed to the content request, as well as,
reduce the network overhead. Compared with the traditional cache
strategies, the proposed NVCP has a great improvement of the
average hop count and average transmission latency. Compared with
LCE, prob(0.5) and MPC, the average hop count of NVCP is reduced
by $0.08\sim0.17$ hops and the average transmission latency is
reduced by $8\sim15ms$.

\section{Concluding Remarks}

By seamlessly converging 5G technologies and telemedicine to
realize the remote surgery, remote consultation and patient
monitoring, people in remote areas can receive high quality
services from developed areas, improving the utilization
efficiency of medical resources and reducing the time and cost of
the diagnosis. In this article, we first characterized the general
architecture of the remote e-health, and then introduced 5G
technologies supported telemedicine to satisfy the high-speed
transmission of massive multimedia medical data, and further
realize the sharing of medical resources. In addition, the BAEPC
scheme was proposed to track and control the spread of the
COVID-19. The challenges, opportunities, and future research
trends, as well as the open issues for the remote e-health are
provided. Finally, the NVCP based caching strategy was
investigated to overcome the massive data storage and low-latency
transmission issues. The interesting future research avenues would
be that introduce the ``Big Data $+$ AI" into telemedicine, to
construct the application of AI assisted diagnosis and treatment;
modeling and analyzing the imaging medical data to provide
decision support for medics and improve the medical efficiency and
quality; with the blockchain technology, encrypt the underlying
data to realize the secure and reliable transmission of medical
privacy data.

\end{document}